\documentclass[preprint2]{aastex}

\usepackage{color}

\slugcomment{to be submitted to ApJ}

\shorttitle{Flared disk around HD200775}
\shortauthors{Okamoto et al.}

\begin{document}


\title{
Direct detection of a flared disk around a young massive star HD200775 \\
 and its 10 to 1000AU scale properties$^*$
}

\author{Yoshiko Kataza Okamoto \altaffilmark{1,**},
        Hirokazu Kataza \altaffilmark{2,**},
        M. Honda     \altaffilmark{3},
        H. Fujiwara  \altaffilmark{4},
        M. Momose    \altaffilmark{1},
        N. Ohashi    \altaffilmark{5},
        T. Fujiyoshi \altaffilmark{6},
        I. Sakon     \altaffilmark{4},
        S. Sako      \altaffilmark{7},
        T. Yamashita \altaffilmark{8},
        T. Miyata    \altaffilmark{7},
     and T. Onaka    \altaffilmark{4},
}

\altaffiltext{1}{
  Institute of Astrophysics and Planetary Sciences, College of Science, 
  Ibaraki University, 
  2-1-1 Bunkyo, Mito, Ibaraki, 310-8512, Japan}
\altaffiltext{2}{
  Department of Infrared Astrophysics, Institute of Space and Astronautical 
  Science, Japan Aerospace Exploration Agency, 
  3-1-1 Yoshinodai, Sagamihara, Kanagawa 229-8510 Japan}
\altaffiltext{3}{ 
  Department of Information Science, Kanagawa University, 
  2946 Tsuchiya, Hiratsuka, Kanagawa, 259-1293, Japan}
\altaffiltext{4}{ 
  Department of Astronomy, Graduate School of Science, University of Tokyo, 
  7-3-1 Hongo, Bunkyo-ku, Tokyo 113-0033 Japan}
\altaffiltext{5}{
  Institute of Astronomy and Astrophysics, Academia Sinica, 
  P. O. Box 23-141, Taipei 106, Taiwan, China}
\altaffiltext{6}{
  Subaru Telescope, National Astronomical Observatory of Japan, 
  650 North A'ohoku Place, Hilo, HI96720, U.S.A.}
\altaffiltext{7}{
  Institute of Astronomy, School of Science, University of Tokyo, 
  2-21-1 Osawa, Mitaka, Tokyo 181-0015 Japan}
\altaffiltext{8}{
  ELT Project, National Astronomical Observatory of Japan, 
  2-21-1 Osawa, Mitaka, Tokyo 181-8588 Japan}
\altaffiltext{*}{
  Based on data collected at Subaru Telescope, which is operated by the 
  National Astronomical Observatory of Japan.}
\altaffiltext{**}{
  These authors contributed equally to this work. 
  Correspondence should be addressed to Y. K. O. (yokamoto@mx.ibaraki.ac.jp).}

\begin{abstract}

We made mid-infrared observations of the 10\,M$_\odot$ Herbig Be star
 HD200775 with the Cooled Mid-Infrared Camera and Spectrometer (COMICS)
on the 8.2\,m Subaru Telescope. We discovered 
diffuse emission of an elliptical shape extended in the north-south direction in $\sim$1000\,AU radius
around unresolved excess emission.
The diffuse emission is perpendicular to the cavity wall formed by the 
past outflow activity and is parallel to the projected major axis of the central close 
binary orbit. The centers of the ellipse contours of the diffuse emission 
are shifted from the stellar position and the amount of the shift increases as the contour 
brightness level decreases. The diffuse emission is well explained in all 
of geometry (the shape and the shift),
size, and configuration by an inclined flared disk where only 
its surface emits the mid-infrared photons. Our results give the first 
well-resolved infrared disk images around a massive star and strongly 
support that HD200775 is formed through the disk accretion. The disk survives 
the main accretion phase and shows a structure similar to that around 
lower-mass stars with 'disk atmosphere'. At the same time, the disk 
also shows properties characteristic to massive stars such as 
photoevaporation traced by the 3.4\,mm free-free emission 
and unusual silicate emission with a peak at 9.2\,$\mu$m, 
which is shorter than that of many astronomical 
objects. It provides a good place to compare the disk properties 
between massive and lower-mass stars.

\end{abstract}


\section{Introduction}

In these two decades, many circumstellar disks around young
forming/formed stars less than several solar masses are found by direct images
in the infrared/visible observations
(e.g. McCaughrean \& O'dell 1996; Fukagawa et al. 2004; Fujiwara et al. 2006;
Lagage et al. 2006; see also web database
\footnote{A comprehensive list of spatially resolved disks
is available (www.circumstellardisks.org).}).
Radio, infrared, and visible line observations
confirmed that some disks have rotating disk
kinematics as expected (Simon, Dutrey, \& Guilloteau 2000;
Pontoppidan et al. 2008; Acke, van den Ancker, Dullemond 2005).
In contrast, the formation scenario for massive stars ($>$8M$_\sun$)
is still unclear. Since, for
such massive stars, the time scale for the Kelvin-Helmholtz
contraction is shorter than that of free-fall or
accretion with accretion rate similar to low mass
star cases, they start releasing energy through nuclear
fusion even during accretion (Palla \& Stahler 1993).
Then radiation pressure due to their large luminosities
may prevent surrounding material from accreting onto the star,
in particular, in the case of very massive stars
(Kahn 1974; Wolfire \& Cassinelli 1987).
Several ideas are proposed to overcome the problem:
mass accretion through circumstellar disks (Yorke \& Sonnhalter 2002;
Krumholz et al. 2009),
mass accretion under a much larger accretion rate than that
usually considered for low mass stars (McKee \& Tan 2002;
Krumholz et al. 2009),
or merging of low mass stars (Bonnell, Vine, \& Bate 2004).
Among these ideas, non-isotropic accretion
through their circumstellar disks seems most plausible at present.
Such non-isotropic accretion alleviates effective radiation
pressure on the accreting material. Supporting evidence
for this disk scenario is recent discoveries of rotating gas
fragments around possible massive young stellar objects (YSOs)
by interferometric observations in the radio,
especially in millimeter and submillimeter
wavelength regions (Cesaroni et al. 2007).
Some of them have a velocity gradient in a direction orthogonal to
molecular outflow lobes, which suggests that the gas fragments
rotate around the central YSOs (Zhang, Hunter, Sridharan 1998; 
Cesaroni et al. 2005).
About a dozen of objects as candidate disks around massive YSOs 
of up to 20\,M$_\sun$ are found so far 
(Zhang et al. 1998; Patel et al. 2005; Cesaroni et al. 2006; 
Beltr\'{a}n et al. 2006; Cesaroni et al. 2007).
Typically, their 
estimated stellar mass, luminosity, and disk radius are 4 to 
less than about 20\,M$_\odot$, a few $\times$(10$^3$--10$^4$)\,L$_\odot$,
and 500--2000\,AU, respectively.
It is suggested that early B Herbig Be stars are surrounded by
flattened structures from measurement of depolarization across 
H$\alpha$ line although the discussed scale is much smaller 
(order of several stellar radii) than the disk size indicated above
(Vink et al. 2002).

While radio interferometric observations have so far been the most 
successful in unveiling the disk existence around massive 
YSOs, the resolution around 1$''$ is not sufficient to draw the 
detailed disk geometry. It is in contrast to the situation that disks 
around young forming stars less than several solar masses have been well 
depicted by direct images in visible to infrared wavelengths.
For some massive YSOs, existence of disks is discussed from
the polarization vector distribution of infrared scattered light
image of the outflow cavities (Jiang et al. 2005), but their disks
themselves are not seen because such objects are still embedded
deeply in their envelopes. Direct images are strongly required to
establish the existence and shape of the disks around massive YSOs.
 Our new approach is to 
search for disks in the mid-infrared around massive YSOs that have emerged 
from their natal clouds. Owing to the large luminosities of the central 
stars, the disk surface can be heated up out to large radii enough to 
be resolved in the mid-infrared with 8\,m-class telescopes, which 
provide 100\,AU resolution for nearby ($\sim$400\,pc) targets.
We carry out survey 
observations for extended emission around Herbig Be stars and report 
the discovery of a disk around HD200775 in this paper.

\section{HD200775}

HD200775 is a Herbig Be star located at 430$^{+160}_{-90}$\,pc
 from the Sun
\footnote{
 The distance, 430pc, is derived by observations with the 
Hipparcos satellite (van den Ancker et al. 1997) based on the 
parallax measurement. Monnier et al. (2006) re-examined the 
distance taking account of the effect of the close-binary orbit 
and estimated it to be 360$^{+120}_{-70}$pc. Their estimate, however, 
also suggests a much lower stellar mass for HD200775, which 
obviously conflicts with the established spectral type. 
Since the original estimate of 430pc agrees with the 
distance by Monnier et al. (2006) within the error, 
we adopt 430pc as the distance in this paper.
} (van den Ancker et al. 1997). The spectral type is estimated to be
B3($\pm$1)e based on the equivalent width of several optical lines 
such as \ion{He}{1}+\ion{Fe}{1} at 4922\,\AA \, and H$\delta$ at 4102\,\AA 
\,(Hern\'{a}ndez et al. 2004).
Its intrinsic bolometric luminosity is estimated as 5400 to 15000\,L$_\odot$ 
after corrected for the extinction by assuming $R_\mathrm{v}$=3.1 to 5.0, 
respectively (Hern\'{a}ndez et al. 2004), where $R_\mathrm{v}$=5.0 better agrees
with the extinction derived independently from B-V and V-R colors.
HD200775 appears above the zero-age main sequence 
on the HR diagram (Hern\'{a}ndez et al. 2004), which supports the youth
of the system as does the 
CO biconical cavity formed by the outflow activity extending 
in the east-west direction with a size of 1.5\,pc$\times$0.8\,pc 
(Fuente et al. 1998a; Ridge et al. 2003). 
No high velocity gas component related to the outflow is currently seen,
which suggests that the system is 
at a stage after its outflow activity.
The observed CO gas wall corresponds to the ridge of the famous 
optical reflection nebula NGC7023 (Fuente et al. 1992), which is 
illuminated by HD200775.
Since the gas wall around the east and west cavities has red- and 
blue-shifted velocity components, the outflow is thought to be 
observed as an almost edge-on configuration (Fuente et al. 1998a).

The stellar mass of HD200775 is estimated by various studies.
Hern\'{a}ndez et al. (2004) show that HD200775 is located at
the evolutionary tracks of a star with 8.9 to 12.5\,M$_\odot$ on the
HR diagram, which
evolvs into a B3--B0 zero age main sequence star.
Based on the time variations of the H$\alpha$ line,
Miroshnichenko et al. (1998) point out that the star might have a
close companion.
Recent near-infrared interferometric 
multi-epoch observations derive the orbital parameters of the close binary,
such that the semi-major axis is 15.14$\pm$0.70\,mas
(corresponding to 6.5\,AU at 430\,pc) and the orbital period is
1377$\pm$25days (Monnier et al. 2006).
On the basis of these data, Monnier et al. (2006)
independently estimated the mass of the sum of the primary ($M_p$) and
companion stars ($M_s$) as 
$M_p+M_s$=10.4$^{+20.5}_{-5.9}$\,M$_\sun$
\footnote{
Monnier et al. (2006) used an orbital period of $P$=3.74$\pm$0.06\,yr,
which is the weighted average between their astrometric measurement and radial
velocity measurement by Pogodin et al. (2004). They assumed the distance
as 430$^{+160}_{-90}$\,pc.
}.
The mass ratio of the 
companion to the whole stellar system is derived as 0.175$\pm$0.035 
from the variability and the bisector radial velocity of H$\alpha$ emission
of the system (Pogodin et al. 2004). 
All these characteristics indicate that the 
primary star of HD200775 is likely to be a massive YSO with the mass 
larger than 8\,M$_\odot$ or more.

Alecian et al. (2008) made monitoring observations of the polarization
in H$\alpha$ and some photospheric lines of HD200775. 
They fit a number of photospheric absorption lines with a combination of
the the primary and the secondary stars and estimate 
the orbital parameters. They obtain $P=$3.87$\pm$0.15\,yr, which is similar
to those obtained by Pogodin et al. (2004) and Monnier et al. (2006).
They attribute shaper, deeper absorption lines to the primary and broader, 
shallower absorption lines to the secondary star.
They find that the radial velocity of the H$\alpha$ line follows that 
of the secondary
\footnote{
From the time variation of the radial velocity of H$\alpha$,
Alecian et al. (2008) derived the mass ratio $M_p/M_s$=0.81$\pm$0.22.
Note that the assignment of the primary and the secondary 
by Alecian et al. (2008) is based not on the measured mass but on the line 
profiles. As a result, their analysis indicates that the primary star
has a larger velocity variation and thus the primary star has a less
mass than the secondary.
On the other hand, they estimated the mass ratio as $M_p/M_s$=1.1$\pm$0.5 
from the luminosity analysis described below, which is consistent with the
orbital analysis.
Both of Alecian et al. (2008) and Pogodin et al. (2004) show the time
variation plot of the H$\alpha$ radial velocities, where
that assigned as the secondary's one by Alecian et al. (2008) is
very similar to the bisector velocity by Pogodin et al. (2004) who
concluded it to follow the velocity of the primary.
See Alecian et al. (2008) for the details.
}.
All lines observed in their spectrum of the secondary are also detected in the 
spectrum of the primary. They speculate that the higher rotational velocity
and slightly fainter luminosity of the secondary star make the lines broader 
and shallower, concluding that the temperature of the secondary star is 
similar to that of the primary star. The optical spectrum of HD200775 can be 
fitted with a combination of two stars better than with a single star.
From the fitting, they estimated that the primary and the secondary stars 
have similar effective temperatures (18600$\pm$2000\,K) and that luminosity
ratio as $L_s/L_p$=0.67$\pm$0.05. With the total luminosity of
15000L$\sun$ (Hern\'{a}ndez et al. 2004) and the luminosity ratio, the two
stars can be plotted on the HR diagram independently, 
from which the stellar masses are
estimated as 10.7$\pm$2.5 and 9.3$\pm$2.1\,M$_\sun$, respectively.
The total mass (20.0M$_\sun$) agrees with a simple estimate of 
the total mass $M_p+M_s=4\pi^2a^3/(GP^2)=19.9$M$_\sun$, which is
directly derived from the orbital parameters by Monnier et al. (2006)
and the distance of 430\,pc
\footnote{The value is different from Monnier et al. (2006), but
is still within their errors that take account of uncertainties in
several parameters systematically.
}.
The orbital parameters, luminosity, spectrum analysis, and spectral type 
can all be accounted for consistently by the assumed distance of 430\,pc.
Thus, it is most likely that the HD200775 system has at least one massive
star of $\sim$10M$\sun$ and that the 15000\,L$\sun$ is close to the 
real total luminosity.

From the location on the primary and
secondary stars on the HR diagram, Alecian et al. (2008) estimated the system age is
0.1$\pm$0.05\,Myr from the birthline. It is in contrast with
Fuente et al. (1998b), who suggest it more likely that HD200775 is a 
post-main sequence star that left the main sequence 8\,Myr ago,
just because a pre-main sequence star has a very short life time.
Alecian et al. (2008) also
estimated the inclination of the binary orbit and of the stellar
rotation axis of the primary star as 48$^{+17}_{-13}$\,degrees and
60$\pm$11\,degrees, respectively. The latter one is derived from 
the analysis of the magnetic field of the stars.

\section{Observations and Reduction}

\subsection{Mid-Infrared Imaging with the COMICS}

The imaging of HD200775 with the Cooled Mid-Infrared Camera and 
Spectrometer (COMICS; Kataza et al. 2000; Okamoto et al. 2003; 
Sako et al 2003; Ishihara et al. 2006) on the 8.2\,m Subaru Telescope 
was made with four filter bands at 8.75 ($\Delta\lambda$=0.752)\,$\mu$m,
11.74 ($\Delta\lambda$=1.058)\,$\mu$m, 18.75 ($\Delta\lambda$=0.90)\,$\mu$m,
and 24.56 ($\Delta\lambda$=0.75)\,$\mu$m. 
The observations were made on 2006 August 20-21 and 2007 June 5-6 (UT) 
for the 8.8 and 11.7\,$\mu$m bands, and on 2008 July 19 (UT) for the 18.8 and 
24.5\,$\mu$m bands. The pixel scale in the imaging is 0.13$''$/pix and the 
achieved spatial resolutions are 0.26, 0.32, and 0.55, 0.66$''$ 
as the full width at half maximum of the 8.8, 11.7, 18.8, and 24.5\,$\mu$m 
point spread functions (PSFs), respectively. During the observations, 
the secondary mirror was chopped to cancel the background radiation. 
For the 8.8 and 11.7\,$\mu$m observations, 
we used an on-chip chopping method with a 0.43--0.45\,Hz 
and 12$^{\prime\prime}$ chop throw nearly along the disk major axis. 
For the 18.8 and 24.5\,$\mu$m observations,
 we used 20 or 30$^{\prime\prime}$ throw with 0.43\,Hz along the disk minor axis. 
For data reduction, we applied a shift-and-add method to produce the images
as described below.
For the sky subtraction, we used a standard chop subtraction method. 
Flat fielding was done with self-sky flat. The flux calibration 
was made based on Cohen et al. (1999).

For both of the 8.8 and 11.7\,$\mu$m filter bands, we repeated many short 
exposures (100\,ms and 80\,ms, respectively) for the object and the standard 
stars (HD198149 or HD186791) as the flux and the PSF reference. 
The standards were observed either before or after HD200775. 
Because the observations were made on several nights, 
we have several sets of the object and standard exposures pairs. 
The numbers of the pairs is 3 and 4, and the total integration 
time of the object observations for each pair is 43--108\,sec and 43--96\,sec, 
for the 8.8 and 11.7\,$\mu$m band, respectively.
For each exposures pair, we divided the original pixel into 
10$\times$10 sub-pixels and applied shift-and-add image stacking to 
keep the spatial resolution of the stacked images. Then we 
obtained seven sets of final images. 
Fluctuations among these sets are taken into account in the 
uncertainty in the disk model fit in section 4.1.
Finally we stacked all the pair images.
The total integration time summed up for all the pairs
amounts to 195 and 243\,sec for the 8.8 and 11.7\,$\mu$m bands, respectively.

For the 18.8\,$\mu$m filter band, the total integration time is 3134\,sec. 
The PSF reference is HD3712 and the flux references are 
HD3712, HD189319, HD186791, HD198149, and HD213310. 
Reduction similar to the 10\,$\mu$m images was made.

For the 24.5\,$\mu$m filter band, the total integration time is 4412\,sec. 
The flux and PSF references are HD3712, HD189319, and HD213310. 
We employed a shift-and-add method, but we did not divide the 
original COMICS pixels because the object was faint at 24.5\,$\mu$m.
Nevertheless, accuracy of the peak detection was good enough (0.12$^{\prime\prime}$ in rms).

The obtained images are shown in Figures 1 and 2. The brightness profiles at 
the 10\,$\mu$m bands are shown in Figure 3.

\subsection{Spectroscopy with the COMICS}

We have made low dispersion (R$\sim$250) spectroscopy at the 10\,$\mu$m 
region with the COMICS on the Subaru Telescope on 2006 August 20--21 (UT). 
We used the long-slit of 0.33$^{\prime\prime}$$\times$36$^{\prime\prime}$ 
in size along the disk major axis 
and 0.45\,Hz chop with 12$^{\prime\prime}$ throw along the slit. We measured the slit 
position precisely from the slit-viewer images. For the flat-fielding, 
we used thermal spectra of the cell-cover of the primary mirror.
The distortion of the spectra 
on the detector was corrected based on the standard stars, which 
were observed with the same instrument rotator angle as used for 
HD200775. The wavelength was calibrated based on the atmospheric 
emission lines. The atmospheric absorption was corrected by the 
spectra of the standard star HD198149. The spectra are calibrated 
to give the absolute flux and the wavelength dependence of the slit 
efficiency. The standard deviation of the sky 
background in each wavelength is taken as an error. The obtained 
spectra at several positions are shown in Figure 4.

\subsection{ 350.3GHz Observations with the Submillimeter Array}

Submillimeter Array
\footnote{
The Submillimeter Array is a joint project between the 
Smithsonian Astrophysical Observatory and the Academia Sinica 
Institute of Astronomy and Astrophysics and is funded by the 
Smithsonian Institution and the Academia Sinica.
}
(SMA; Ho, Moran, \& Lo 2004) observations of the 
continuum emission at 345\,GHz in the lower side band and at 355\,GHz 
in the upper side band were made on 2008 July 9th (UT). 
The extended array configuration consisting of 7 antennas 
provided projected baseline lengths between 33\,k$\lambda$ 
to 212\,k$\lambda$, and 
the largest structure to which these observations are sensitive
enough (more than 50\% level) is expected to be 2.7$''$
(Wilner \& Welch 1994).
3C279 was observed to obtain the passband characteristics of the
system, while the two quasars 1849+670 and 2009+724 were observed
every 15 minutes to track the time variation of the complex gain.
Visibility data were calibrated with IDL MIR package for the SMA.
After the correction for the visibility amplitude by using the
record of system temperature, the gain table constructed from 1849+670
data was applied to the visibilities of HD200775. Flux density of
1849+670  at 350GHz was estimated to be 1.50 Jy from the comparison
with Uranus.
Both the data sets from the upper and lower side bands are combined, 
and natural weighting is applied in the image construction process. 
The synthesized beam size (, or PSF) is 0.87$''$$\times$0.78$''$ (FWHM) 
and its position angle of the major axis is 27.7 degrees. 
The obtained image is shown in Figure 5.

\section{Results and Discussion}

\subsection{The Mid-Infrared Images and a Flared Disk Model}

  The obtained 8.8, 11.7, and 18.8\,$\mu$m images (Figs. 1 and 2) have enough 
S/N ratios and consist of unresolved source emission and elliptical 
diffuse emission of 2$''$ in the semi-major axis, whose shape appears as an 
inclined disk. 
At 18.8\,$\mu$m, additional diffuse emission with a curved tail extending 
toward the northeast is also seen. 
The observed total fluxes are 
9.4$\pm$0.2\,Jy, 6.7$\pm$0.7\,Jy, and 26.2$\pm$1.3\,Jy 
at 8.8, 11.7, and 18.8\,$\mu$m, respectively. 
The fluxes of the unresolved component are 5.4\,Jy, 3.5\,Jy, and 2.0\,Jy, 
which are much larger than the expected photospheric emissions 
0.2\,Jy, 0.1\,Jy, 0.04\,Jy, at 8.8, 11.7, and 18.8\,$\mu$m
 (Acke \& van den Ancker 2004), respectively. 
The size of unresolved source emission is estimated from 
comparison of the azimuthally averaged radial profiles 
between the unresolved source and the PSFs. If we assume a 
Gaussian radial profile for the unresolved source, it is 
estimated that the 95\% of the unresolved flux arises from 
$r<$22AU and $r<$29\,AU at the 8.8 and 11.7\,$\mu$m bands, 
respectively.
 Thus the unresolved mid-infrared emission arises mostly from 
the circumstellar dust in the vicinity of the central star(s).
The elliptical emission has a position angle of 6.9$\pm$1.5 degrees 
and is detected out to 750--1000\,AU in radius along the major axis (Fig. 3). 
It is similar to the radii of disk candidates around massive YSOs 
(Cesaroni et al. 2007) and disks around low and intermediate mass stars 
(200--1000\,AU; Dutrey, Guilloteau, \& Ho 2007), while it is much 
smaller than the envelope radius (28000\,AU) estimated from the north-south 
distance of the CO cavity wall at the apex. The major axis of the 
diffuse elliptical emission is orthogonal to the cavity. 
Also the major axis is in the same direction as the ascending node 
(PA= -0.2$\pm$7.6 degree) of the binary orbit.
All the observed shape, size, and 
configuration support that the elliptical emission arises from the 
circumbinary disk around HD200775, which is in almost the same plane
as the binary orbit. The results strongly support
that the star formed through the disk accretion and that the 
disk around a massive star still remains even after the outflow 
phase, that is, the main accretion phase. The 18.8\,$\mu$m tail extends 
toward the southern edge of the northern part of the CO cavity apex. 
It may indicate a structure that bridges the disk and the outer region. 
The 24.5\,$\mu$m image is relatively similar to that at 18.8\,$\mu$m.

For the disk emission, the isophotal contours are roughly elliptical. 
The ellipse center is shifted toward the east from the unresolved 
flux peak source, where the amount of the shift is larger for the fainter isophotal 
contours. 
Brightness contribution from the PSF skirt of the unresolved 
source is still not negligible (cf. figure 3), especially for
the inner radius region. 
However, the shift phenomenon cannot be explained only by
the diffraction effect.
Such a shift is characteristic to a flared disk, 
where the ratio of disk height $z(r)$ to radius $r$ increases 
monotonically toward the outer region and only the surface 
emits the mid-infrared photons (Figure 6). It is consistent with 
that HD200775 shows a spectral energy distribution (SED) steeply 
arising toward the far-infrared (Acke and van den Ancker 2004), 
suggesting a circumstellar flared disk whose surface is irradiated 
by the central star even at outer regions (group I by Meeus et al. 2001). 
It is very similar to the geometry confirmed for the intermediate 
mass young star HD97048 (Herbig B9.5e/A0 star with 2.5\,M$_\sun$ 
and 40\,L$_\sun$; Lagage et al. 2006).
In the case of HD200775, the east side surface faces us.

To derive the disk geometry, we employed a model image fitting method.
We assume a simple axisymmetric disk model, where the emitting 
surface is flared as $z(r)=z_0 (r/r_0)^{\alpha} $ 
from the inner edge ($r_{in}$) to 
the outer edge ($r_{out}$), to see if the flared disk accounts for our 
obtained images.
 It is assumed that only the disk surface emits in the mid infrared,
that the disk surface beyond the 
midplane is completely extinguished, and that the surface 
brightness of one side per projected midplane area is given by 
$F(r)=F_0 (r/r_0)^{\beta}$.
It corresponds to a passive disk where the disk is mainly heated 
by the radiation onto the disk surface, which is consistent with
the fact that the outflow is not active in HD200775's case.
The unresolved source is not included in the model, but subtracted 
from the observed images by a 'clean' analysis method. The model 
disk is inclined toward us by an angle $i$ from face-on and is 
convolved with the PSF. It is fitted to the observed (peak-subtracted) 
HD200775 images at 8.8 and 11.7$\mu$m with free parameters of 
$r_{in}$, $r_{out}$, $\alpha$, $\beta$, $z_0$, and $F_0$. 
We use individual exposure pairs for the fitting because we 
cannot obtain good PSF images for the pair-averaged images. 
The best fit result (Figure 1i) shows that the observed image 
and the shift in the centers of the contours are well reproduced 
by the simple model with 
$r_{out}$=679$\pm$48\,AU,
$r_{in}$=159$\pm$25\,AU, 
$\beta$=-1.8$\pm$0.3, 
$i$=54.5$\pm$1.2\,degrees 
and $z(r)$[AU]=(19.6$\pm$16.9) $\times$($r$[AU]/280\,AU)$^{2.1\pm1.1}$.
For each observed pair image, the difference between the model 
and the observed images is less than twice the sky noise level
on the average
\footnote{
In detail, we combined 5$\times$5 subpixels into one pixel at the fitting
to reduce calculation time. The difference between the observed image after
the clean and the model diffuse disk image is calculated, is squared, and is 
summed up for the fitted region. The value averaged for the combined pixels is
minimized by the fitting. The resulted values for individual pair images
are 1.1 to 1.8 times larger than the spatial deviation of the sky area of the 
observed images. We did not calculate the $\chi^2$, because the pixel flux
of the observed images are not independent because of the pixel size is
smaller than the PSF sizes.
}
.
The inclination of the disk agrees with the inclinations of the
binary orbit and the stellar rotation axis derived by Alecian et al. (2008;
see $\S$2)
within errors, suggesting that the axes of the stellar rotation, 
binary orbit, and the disk are aligned relatively well.
The observed brightness profile (fig. 2a) is proportional to 
$r^{-1.0\sbond-1.8}$ at $r\lesssim$500\,AU, 
where the agreement with our disk model is fairly good. 
In the outer region, the observed
profile becomes steeper as  $r^{-2.5\sbond-2.6}$ in the south
or more in the north. 
The fitted outer radius is similar to the radius where the slope of
the observed profile becomes much steeper  in the north
($\propto r^{-4.5}$ in the 11.7\,$\mu$m band
 and $\propto r^{-8.7}$ in the 8.8\,$\mu$m band).
Most of the fit parameters do not show significant difference between
8.8 and 11.7\,$\mu$m, while the fitted beta
is slightly larger for 8.8\,$\mu$m (2.0$\pm$0.2) than for 11.7\,$\mu$m
(1.65$\pm$0.24). It seems to correspond to the shallower brightness slope for
11.7\,$\mu$m and suggests that the fit procedure works well.
The real inner radius could be smaller due to possible 
false removal of the emission in the clean procedure, where we 
allowed multiple peak detection within 0.33$''$ radius for 8.8\,$\mu$m
and 0.46$''$ for 11.7\,$\mu$m from the brightest peak. 
The criteria are adopted since very short-time seeing fluctuation 
cannot be treated perfectly even in the shift-and-add process and 
could produce spurious sub-peaks. While most detected sub-peaks 
distribute very close to the brightest peak, some sub-peaks are seen 
along the major axis. Part of the emission coming from the inner disk 
could have been removed by this process. In general, the simple flared disk 
model accounts for most of the observed characteristics of HD200775. 
Our results, the first well-resolved direct infrared images around a 
massive star, clearly reveal the geometry of the disk around the 
massive YSO.

The derived flared geometry of the HD200775 disk is much flatter than
the disk around the intermediate mass star HD97048. The maximal and
average opening angle of the HD200775 disk is much smaller than the
HD97048 disk where
$z_{HD97048}(r)$[AU]=51.3$^{+0.7}_{-3.3}\times$($r$[AU]/135\,AU)$^{1.26\pm0.05}$
(Lagage et al. 2006).
Then $z(r)/r$ becomes 0.12 (at 680\,AU radius) to 0.27 (if we extend
our fit result to 1000\,AU radius)
for HD200775 while $z_{HD97048}(r)/r$ is 0.49 (at 370\,AU radius
which is the outer limit discussed by Lagage et al. (2006)).
In fact, it is suggested that disks around massive stars generally
appear to be flatter since $L_\mathrm{ex}/L_\mathrm{*}$ estimated from
the SED is typically much smaller for early Herbig Be stars than in Herbig Ae stars
(e.g. Acke \& van den Ancker 2004).
It supports the idea that there is a qualitative difference between
disks around massive stars and low mass stars.
It may be related to rapid disruption of disks around massive stars
due to the large luminosities and/or photoevaporation (see $\S$4.4).

In a hydrostatic disk in Keplerian rotation around a star,
it is expected that $\alpha$ becomes $<1.5$ (e.g. Dutrey et al. 2007).
Here we assume that the hydrostatic structure
is determined based on the temperature at the equatorial plane and
the surface density of the disk, both of which have power-law dependence on radius.
Also we assume that the temperature decreases as the radius.
Owing to the large uncertainty in $\alpha$, we could not conclude 
from the model fit whether or not the HD200775 disk is flared more 
steeply than the HD97048 disk ($\alpha$=1.26), which is consistent with a 
hydrostatic one.
In our fitting, we fix $\alpha$ between 0 and 4.1 with a step
of 0.1 and search for a combination of the free parameters that minimizes the
squared sum of difference between the model images and the observed images.
We find that $z_0$ is strongly correlated with $\alpha$ in our
model fitting. Thus $z_0$ and $\alpha$ can not be determined independently.
However, the disk height within $r<R_{out}$ tends to become similar
even if the combination of $z_0$ and $\alpha$ changes. 
The other parameters ($R_{in}$,
$i$, $\beta$, $F_0$) does not change largely.

The disk detected here is clearly related with the past
outflow activity, binary orbit, and inclination of the star.
It strongly suggests that this star is still at
a pre-main sequence stage, as discussed by Alecian et al. (2008).
The present observations indicates the presence of a disk 
around a high-mass pre-main sequence star.

\subsection{Thermal Structure of the Circumbinary Disk around HD200775}

Figure 7 shows the distribution of color temperature of the HD200775 
circumbinary disk. It is estimated by using the brightness ratios of 
the 11.7 and 18.8\,$\mu$m images where the unresolved components are 
subtracted. The difference of the diffraction size in the two bands 
is not corrected here. In addition, the 11.7 and 18.8\,$\mu$m bands
contain contributions from the silicate feature emission, from which
the estimated temperature may suffer.
The ISO/SWS spectrum toward HD200775
shows the moderate silicate emission at 10\,$\mu$m region, but
has no prominent silicate emission at 18.8\,$\mu$m region.
The 11.7\,$\mu$m band is located at the skirt of the 10\,$\mu$m silicate feature.
Thus the estimated temperatures may be overestimated.
Since it is not necessary that 
the flux at the two bands comes exactly from the same grains and 
areas, the actual temperature distribution would be more complicated: 
grains at the surface has higher temperatures and those 
at a deeper location in the disk have lower temperatures. 
Figure 7 indicates that the color temperature is $\sim$180--330\,K. Black 
grains becomes only 113\,K or 149\,K at 430\,AU 
and 77\,K or 97\,K at 1000\,AU from a 6000\,L$_\odot$ or 15000\,L$_\odot$ 
star, respectively. It suggests that the grains are not blackbodies
but they are small and absorb the stellar photons efficiently in
visible wavelengths and radiate in the infrared inefficiently. 
Figure 8 shows the temperature distribution of blackbody and small 
(radius $a$=0.1 and 1\,$\mu$m) grains that absorb
the stellar light completely but have an emissivity of $(a/\lambda )$
in the infrared.
It suggests that grains of $\sim$0.1\,$\mu$m radius have 
temperatures similar to that of the observed diffuse disk emission.

By assuming amorphous Mg-pure pyroxene grains with 0.1\,$\mu$m radius 
(J\"{a}ger et al. 2003;  see section 4.5) at 300\,K, we derive 
the mass of the dust emitting at 10\,$\mu$m as 8$\times 10^{-9}$\,M$_\sun$. 
The total mass including gas is very small (8$\times 10^{-7}$\,M$_\sun$),
when the gas-to-dust mass ratio is assumed as 100. 
Also it is difficult to heat up grains in the midplane warm enough 
($\gtrsim$300\,K) to be detected in the 10\,$\mu$m bands at the observed large 
radii in a passive disk. We conclude that the mid-infrared disk 
emission arises from the disk atmosphere, or the superheated surface 
layer illuminated directly by the central star as modeled for 
low to intermediate mass stars (Chiang \& Goldreich 1997, Chiang et al. 2001, Dullemond \& Dominik 2004). 
If the beam filling factor by the diffuse disk emission is assumed as 
unity, the brightness at 10 and 18.8\,$\mu$m suggests the 
optical depth in the mid-infrared ($\tau_\mathrm{MIR}$) of 
$10^{-4}$--$10^{-5}$ and $10^{-3}$--$10^{-4}$ 
for the dust temperatures of 300\,K and 200\,K, respectively. 
It corresponds to the optical depth at 0.55\,$\mu$m ($\tau_\mathrm{v}$) of
$10^{-2}$--$10^{-4}$.  It should be noted that the optical depth 
$\tau_\mathrm{v}$ normal to the disk surface is smaller than most model predictions
(e.g. Chiang \& Goldreich 1997; $\sim z(r)/r \sim 0.1$ for the HD200775 disk)
where it is assumed that the surface height is proportional 
to the vertical gas scale height at any radius and that the 
absorbed flux of the stellar radiation at the disk surface is proportional to 
the sine of the local grazing angle ($F\propto \sin\theta$). 

There are several possible reasons for the small optical depth.
Firstly, the temperatures of the dust emitting in the mid-infrared region
might be lower than those assumed (300K for 10$\mu$m region and
200K for 20$\mu$m region, respectively). As we mentioned, the silicate
features may cause overestimate of the temperatures. If the dust temperature
is 150K, the optical depth in the 10$\mu$m region becomes 60 to 240 times
higher. 
Secondly, the photospheric emission of HD2000775 ($T_*$=19000\,K) peaks
around 0.1$\mu$m in the ultra-violet (UV) region.
The interstellar extinction at 0.1$\mu$m is larger than at the V-band
as $A_{0.1\mu m}$/$A_\mathrm{v} $ is about 5 in a typical case or 
higher ($\sim$10) in other cases (Ryter 1996).
Since the thickness or the optical depth of the disk atmosphere
is determined so that the most stellar emission is absorbed by the disk 
surface, the thickness of the HD200775 disk case is likely to be determined by 
the optical depth at the UV rather than visible wavelengths.
Thirdly, the ratio of extinction between the mid-infrared and optical 
wavelengths for grains in the HD200775 disk may not be the same for 
interstellar grains we assumed (Rieke \& Lebofsky 1985).
In fact, it is suggested that the grains in the HD200775 disk
are different from those in the interstellar medium (ISM) as discussed
in $\S$4.5. If $A_\mathrm{v}/A_\mathrm{MIR}$ in HD200775 is much larger than the ISM,
we will have a larger $\tau_\mathrm{v}$.
The ratio $A_\mathrm{v}/A_\mathrm{MIR}$ becomes larger if the grains 
are much blacker in the
visible to UV wavelengths than in the ISM, or if the grains
have much lower emissivity in the mid-infrared than in the ISM.
Lastly, the small optical depth in HD200775
might be understood in terms of the thermal instability of protoplanetary 
disks. Watanabe \& Lin 2008 predict that thermal wave can be excited to make 
ripples of the disk surface in a disk with a low accretion rate.
They predict that $\tau_\mathrm{v}$ normal to the disk is $\sim$1 only at the
excited ripples (thermal wave pulse) and becomes much lower (a few tenths to 
less than $10^{-3}$) at the other most areas.

The ratio of the infrared excess emission ($L_\mathrm{ex}$)
to the total luminosity ($L_\mathrm{tot}$) is estimated to be 0.07 from the SED
(Acke \& van den Ancker 2004).
The flared geometry derived by our disk model fit gives the
fraction of the intercepted stellar radiation of 0.12--0.27.
Taking account of the radial optical opacity of the surface layer
($\sim \tau_\mathrm{v} / (z(r)/r)$), $L_\mathrm{ex}/L_\mathrm{tot}$
becomes much lower than that from the SED. It can be attributed partly to 
the small $\tau_\mathrm{v}$ again, and partly to the following fact.
In HD200775, extended far infrared emission that roughly
traces the outflow cavity wall (NGC7023) is detected (Kirk et al. 2009;
we also confirmed it with data of AKARI satellite (Murakami et al. 2007)).
Therefore the far infrared flux of the SED (IRAS data and
even ISO data) has a significant contribution from the wall. The
large $L_\mathrm{ex}/L_\mathrm{tot}$ of 0.07
indicates that the cavity wall intercepts a large fraction of the stellar
flux. The disk itself does not have to absorb a large fraction.

\subsection{The 350.3\,GHz Image and Mass Estimate}

From our 350.3\,GHz SMA observations, we detected an unresolved emission 
source of 35.1$\pm$5.2\,mJy at the star position with the 1$\sigma$ noise 
level of 3.0\,mJy/beam (Fig. 5). If we assume the mass absorption coefficient
in the radio region given by Hildebrand (1983) and the spectrum (flux $F$) 
slope from submillimeter to millimeter
wavelengths as $F_\nu\propto \nu^{4}$, the mass opacity at
350\,GHz becomes 0.0085\,cm$^{2}$\,g$^{-1}$.
Then the observed flux gives a lower limit for the total disk mass 
($M_\mathrm{350GHz}$) of 0.022 or 0.010\,M$_\sun$ for the dust temperature 
at the midplane of 50\,K or 100\,K ($T_\mathrm{d,350GHz}$), respectively.
The assumed spectrum slope is in agreement with the SED in 
Acke and van den Ancker (2004). 
It is also compatible with the 
upper limit of the flux at 1.3mm ($<$6\,mJy) of past bolometer 
observations of a 10$''$ beam (Fuente et al. 2001). 
The slope indicates that a majority of the grains in HD200775 is small 
in size and not fully grown since large grains have a slower wavelength dependence.

The mass above is similar to the estimated minimum mass for the protoplanetary 
disk around the Sun. If it is distributed uniformly in a sphere of the 
beam radius (180\,AU), it amounts to visible extinction of $>$190\,mag
toward the central star, which is much larger than the observed 
one (1.8 and 3.0\,mag under $R_\mathrm{v}$=3.1 and 5.0, respectively;
 Hern\'{a}ndez et al. 2004).
Here, we assume the optical to 
mid-infrared extinction ratio given by Rieke \& Lebofsky (1985) and 
$N(\mathrm{H})/A_\mathrm{v}$ value in Cox (2001). 
Thus the mass must be distributed in a flattened structure, 
which consists of the inner part of the disk.

If we assume the brightness of the 2$\sigma$ detection limit of 
our SMA observations 
(6\,mJy/beam) for the extended emission, the mid-infrared extinction 
becomes at least 2.0\,mag (for 100\,K dust) or 4.4mag (for 50\,K dust) at 
11.7\,$\mu$m even in the outer region. Extinction at 8.8\,$\mu$m is 
1.5 times larger than that at 11.7\,$\mu$m (Rieke \& Lebofsky 1985). 
Extinction at 18.8\,$\mu$m is not well known and we assume that it is 
similar to that at 11.7$\mu$m (Draine \& Lee 1984; McClure 2009). 
Then, the mid-infrared emission from the further side of the disk
should be extinguished mostly by the midplane before reaching us even
in the outer region. The observations of the mid-infrared and 350\,GHz
can be interpreted in terms of the thick disk in the mid-infrared.

\subsection{Photoevaporating Disk}

While the disk shows geometry and structures relatively similar to 
low mass stars as described above, there are clear differences, too. 
Since the primary star is hot enough, its ionizing photons can 
photoionize the disk gas. It leads to disk dispersal due to 
photoevaporation in the region where the sound speed of ionized gas exceeds the 
escape speed against gravity. The inner radius 
(photoevaporation radius $r_{ph}$) is estimated as $\sim$70\,AU for 
a 10\,M$_\sun$ star in a weak stellar wind case (Hollenbach et al. 1994). 
The mid-infrared elliptical emission of HD200775 corresponds to the 
region outside $r_{ph}$. The distribution of the free-free emission
at 3.4\,mm detected toward 
HD200775 (Fuente et al. 2001) resembles the elliptical disk and the 
tail in the mid-infrared well in size and shape. Though it was 
considered to arise from the stellar wind previously (Fuente et al. 2001), 
the 3.4\,mm and mid-infrared emissions are likely to trace the surface of 
the photoevaporating disk (Fig. 6).
When we assume the ionizing flux from a B3 zero-age main sequence star
($10^{43.69}$ photons s$^{-1}$ ; Panagia 1973), the mass loss rate
due to the photoevaporation is estimated as
7.7$\times10^{-6}$\,M$_\sun$\,yr$^{-1}$ (Hollenbach et al. 1994).
It suggests that a mass of 0.01\,M$_\sun$ to 0.1\,M$_\sun$ will be dispersed
within an order of 1000 to 10000\,yrs.
The disk around HD200775 seems to be a short-lived phenomenon.

\subsection{Spectra: Innermost Circumstellar Disk and Characteristic Diffuse Silicate Emission}

The spatially resolved 10$\mu$m band spectra of the circumstellar disks
have much information on the properties of grains and related processes
(e.g., Okamoto et al. 2004).
Fig. 4 shows the spectra of the unresolved source and at a typical position 
in the diffuse disk along the major axis. 

The spectrum of the unresolved source (fig. 4a) is dominated by the 
featureless continuum emission and shows the \ion{H}{1} 
line at 12.37\,$\mu$m 
of n=7$\rightarrow$6 and n=11$\rightarrow$8 transitions
with a line flux of (4.1$\pm0.4)\times10^{-16}$\,W\,m$^{-2}$. 
The innermost disk is well heated deep into the midplane and radiates an 
optically thick blackbody emission. The temperature estimated from the 
spectrum is 1600\,K ($>$1000\,K considering the 1 $\sigma$ error of the flux 
calibration), which is close to the evaporation temperature of 
silicate and graphite. Black grains illuminated by a 15000\,L$_\sun$ star 
become 1600\,K at 3.7\,AU from the star. 
Since it is smaller than the binary orbit, the unresolved 
source emission arises not from the circumbinary disk but
from circumstellar disk(s)
\footnote{
Black grains become 1000\,K at 9.5\,AU from a 15000\,L$_\sun$
source. However, Artymowicz \& Lubow (1994) points out that 
a circumbinary disk 
with typical viscous disk parameters around a binary system
with a semi-major axis $a$ tends to have an inner truncated radius
of 1.8$a$--2.6$a$ for the binary eccentricity from 0 to 0.25.
They also point out that the minimum separation between the component
star and the inner circumbinary disk edge is larger than $a$
for the binary eccentricity less than 0.75.
The HD200775 binary orbit has an eccentricity of 0.30$\pm$0.06
(Monnier et al. 2006). Thus, it is unlikely that 9.5AU is larger
than the inner edge of 
the circumbinary system. It is most likely that the unresolved source
emission originates from circumstellar disk(s).
}.
The surface 
area of the blackbody emission ($7\times10^{-6}$\,arcsec$^2$) indicates 
a dust ring (or inner wall) at 3.7\,AU radius with only a 0.06\,AU width 
(or height).
L\'{e}pine \& Nguyen-Quang-Rieu (1974)
detected an OH maser line emission at 1667\,MHz toward HD200775 and 
concluded that the line profile may be formed in a rotating gaseous 
disk-shaped envelope within 8 stellar radii. The detected \ion{H}{1} line 
emission probably stems from the ionized gas in the dust disk surface 
or the innermost gas disk where the grains are evaporated.

The diffuse emission at 1.3$''$ south from the unresolved source
(fig. 4b) shows the amorphous silicate feature 
with a triangular shape. The peak is located at 
9.2\,$\mu$m which is largely shifted from the 9.5--9.8\,$\mu$m 
amorphous silicate peak often seen toward
the interstellar medium (Kemper, Vriend, \& Tielens 2004), 
evolved mass-losing stars (Dorschner et al. 1995), and 
disks around low to intermediate mass young stars (Honda et al. 2006). 
It is also shorter than the averaged peak of absorption features 
of dust grains in outer envelopes of massive protostars and is more 
compatible with pyroxene ((Mg, Fe)SiO$_3$) than olivine ((Mg, Fe)$_2$SiO$_4$). 
We compare the continuum subtracted emission with the normalized emission 
from amorphous silicate grains at 300\,K based on 
J\"{a}ger et al. (2003) (Mg$_{0.7}$SiO$_{2.7}$) and pyroxene (MgSiO$_3$) 
grains with 0.1 and 1.0\,$\mu$m radii. As can be seen in fig. 4b, 
the peak wavelength decreases as the SiO$_2$ fraction in silicate 
increases (Koike \& Hasegawa 1987; J\"{a}ger et al. 2003). 
It suggests the dominance of pyroxene grains or grains with larger 
SiO$_2$ fraction rather than olivine. 
The observed peak wavelength is better matched with Mg$_{0.7}$SiO$_{2.7}$
 grains, 
but MgSiO$_3$ gives a better fit in the feature width. 
There is a report (Tamanai et al. 2007) that SiO$_2$ grains of 
0.5--\,5$\mu$m diameter have a peak between 8.9--9.5\,$\mu$m. 
However, our spectra do not show a feature around 12.5\,$\mu$m 
characteristic to SiO$_2$. Thus the grains are in form of 
silicate and not pure SiO$_2$.
The presence of a 8.9\,$\mu$m feature in \ion{H}{2} regions is
reported by Peeters et al. (2005), but it seems not to coincide 
with the observed 9.2\,$\mu$m feature. 
Grain properties very close to massive stars, especially in disks, 
may be different from those of disks around low mass stars and even 
their own envelopes.

To understand the characteristic dust properties, the photoevaporation,
that is, the ionizing flux or ionized gas might be an important factor.
There are many reports of experiments of irradiating ions such as H$^{+}$
and He$^{+}$ on cosmic grains or model grains.
Irradiating
crystalline olivine with high energy protons of a few MeV, typical of cosmic rays, makes
no alteration of the crystalline structure at fluences relevant to the ISM (Day 1977).
In contrast, irradiation of H$^+$ and He$^+$ ions of a few to several 
tens keV can amorphize olivine and pyroxene crystals and
change the chemical composition of the grains
(Bradley 1994; Dukes et al. 1999; Demyk et al.2001).
Demyk et al. (2001) made an irradiation experiment on crystalline olivine with 
He$^{+}$ ions at energies of 4 to 10keV and fluences from 5$\times 10^{16}$
to 10$^{18}$ ions\,cm$^{-2}$. It caused amorphyzation of olivine
in conjunction with an increase in the porosity of the material
due to the formation of bubbles. Furthermore, the amorphized layer was deficient
in oxygen and magnesium. They found that the O/Si and Mg/Si ratios decreased
as the He$^{+}$ fluence increased. It suggests the change of the olivine
to pyroxene structure in the grains.
Bradley (1994) investigated a pyroxene crystal on the surface of an 
interplanetary dust particle and its rim, where alternation has occurred
by the irradiation by solar wind ions. 
He found that the rim was stoichiometrically depleted in Mg and Ca and
enriched in S, Si, and Fe relative to pyroxene. He also made irradiation
experiments of 20keV protons onto a sample of fine-grained matrix from
the Allende chondritic meteorite. Irradiation of both of olivine and pyroxene
crytals in Allende again led to the stoichiometric depletion of Mg and Ca.
He concluded that ion irradiation first striped cations such as Mg and Ca with
the weakest bond strengths and then more strongly bound cations such as Si.
These results indicate that irradiation of energetic ions such as 
H$^{+}$ and  He$^{+}$  causes a change in the lattice structure of grains from 
olivine to pyroxenes, then to more SiO$_2$ rich structures.
Since the HD200775 disk is photoevaporating and a stellar wind activity
is suggested (Skinner, Brown, \& Stewart 1993),
the grains of the HD200775 disk might be alternated by such a mechanism 
in plasma of the disk surface or by the stellar wind. 
Further study is required to investigate whether or not the 
alternation is a common phenomenon in the formation of massive stars.

\section{Conclusions}
We have made mid-infrared imaging and spectroscopy of the Herbig B3e star
HD200775 of $\sim$10\,M$_\sun$ with Subaru/COMICS.
We found elongated emission around the unresolved excess emission 
in all of the 8.8, 11.7, 18.8, and 24.5$\mu$m bands. The elongated
elliptical emission has a radius of about 1000\,AU and configuration
perpendicular to the CO cavity wall formed by the past outflow and
parallel to the projected major axis of the central close binary orbit.
The brightness contours of the elliptical emission are shifted from the
stellar position or from the unresolved source position. All of these
observed characteristics are well explained as the surface emission of
the tilted flared circumbinary disk. 
The present observations provide the first well-resolved infrared images of a disk
around 10\,M$_\sun$ YSO and strongly support that HD200775
is formed through the disk accretion.
The optical depth and the silicate features of the 
disk emission support the 'disk atmosphere' configuration, which is
well modeled for low to intermediate mass stars.
We fit the observed image with a flared disk model.
The overall observed structure is well explained by the model.
The derived flared geometry is much flatter than
the disk around the intermediate mass star HD97048. The value
of $z(r)/r$ is estimated to be 0.12 to 0.27 for the HD200775 disk,
while it is 0.49 for the HD97048 disk.
It supports the idea that there is a qualitative difference between
disks around massive stars and lower mass stars.
The 3.4\,mm free-free emission image in literature is very similar to
the mid-infrared elliptical disk emission in size and shape. It suggests
that the disk surface photoevaporates due to the ionizing 
photons from the central star.
The SMA 350\,GHz observations detect the unresolved emission
with a 0.8$''$ beam. The disk mass concentrated
around the star is estimated as around 0.02\,M$_\sun$, which is similar
to that of the minimum mass the solar nebula.
The 10\,$\mu$m region spectra of the peak unresolved
source and at the 1.3$''$ south are discussed.
The spectrum of the unresolved source  
shows 1600\,K featureless emission, which
correspond to the innermost circumstellar disk 
in the vicinity of the central star(s).
The spectrum of the diffuse emission at 1.3$''$ south 
is dominated by the amorphous silicate emission. 
The peak at 9.2\,$\mu$m is shorter than the usual silicate features
and unique to the HD200775 disk. It may suggest alternation of 
grains due to plasma irradiation.

\acknowledgements

We thank H. Nomura for the helpful discussion with her and Y. Doi
for his help in examining the AKARI data of HD200775.
We thank M. Tosa, S. Takita, K. Sato, K. Manabe, and K. Tomita for their
help at the Subaru observations and reduction.
We would like to thank all of the staff members of the Subaru Telescope
and SMA for their support during the observations and development of 
the instrument. 
We are grateful to the anonymous referee for the helpful comments
which significantly improved this paper.
This work was supported by Grant-in-Aid for Scientific
Research on Priority Areas from the Ministry of Education, Culture, Sports,
Science and Technology, Japan. 
Y. K. O. was also supported by Grant-in-Aid for Young Scientists (B)
and (A), by the Ministry of Education, Culture, Sports, Science and Technology, Japan.

\clearpage

\begin{figure}
\epsscale{2.00}
\plotone{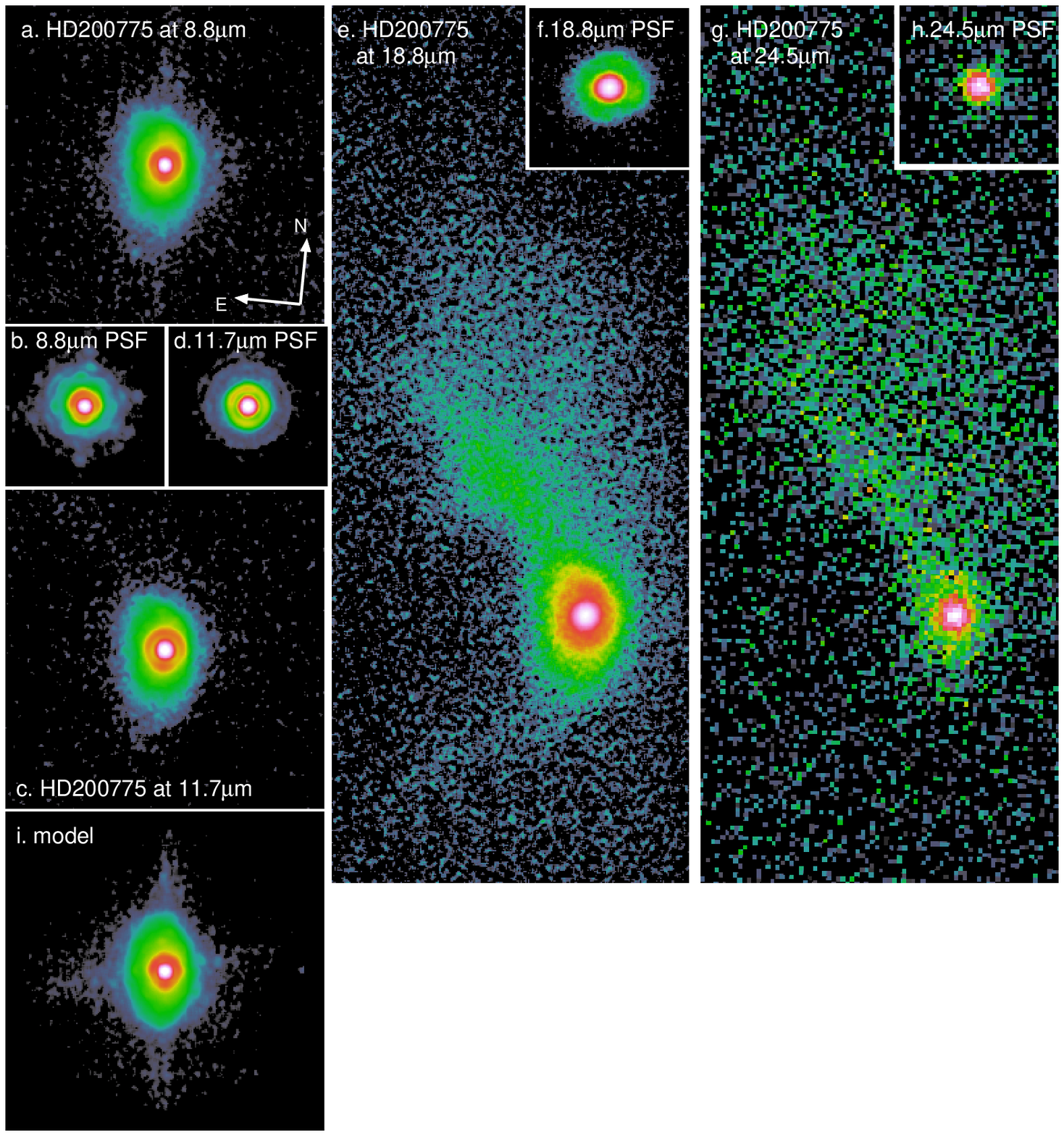}
\caption{
(a--h) The 8.75, 11.74, 18.75 and 24.56\,$\mu$m images of HD200775 
(8.8$''$$\times$8.8$''$ area for the 8.8 and 11.7\,$\mu$m bands; 
 9.9$''$$\times$24.3$''$ area for the 18.8 and 24.5\,$\mu$m bands) 
and the normalized PSF reference images (for 4.4$''$$\times$4.4$''$ area). 
In all figures (a-i), the disk's major axis is along the vertical 
axis and the scale is the same.
The color distribution is assigned from the peak brightness level 
down to the sky noise level for the object images. For the PSF images,
from 1 down to the ratio of sky noise level over the peak brightness
of the object images at the same filter bands.
For the detailed brightness levels, see Figure 2. 
(i) Best fit model image at 8.8\,$\mu$m (see section 4.1).
}
\end{figure}

\clearpage

\begin{figure}
\epsscale{1.00}
\plotone{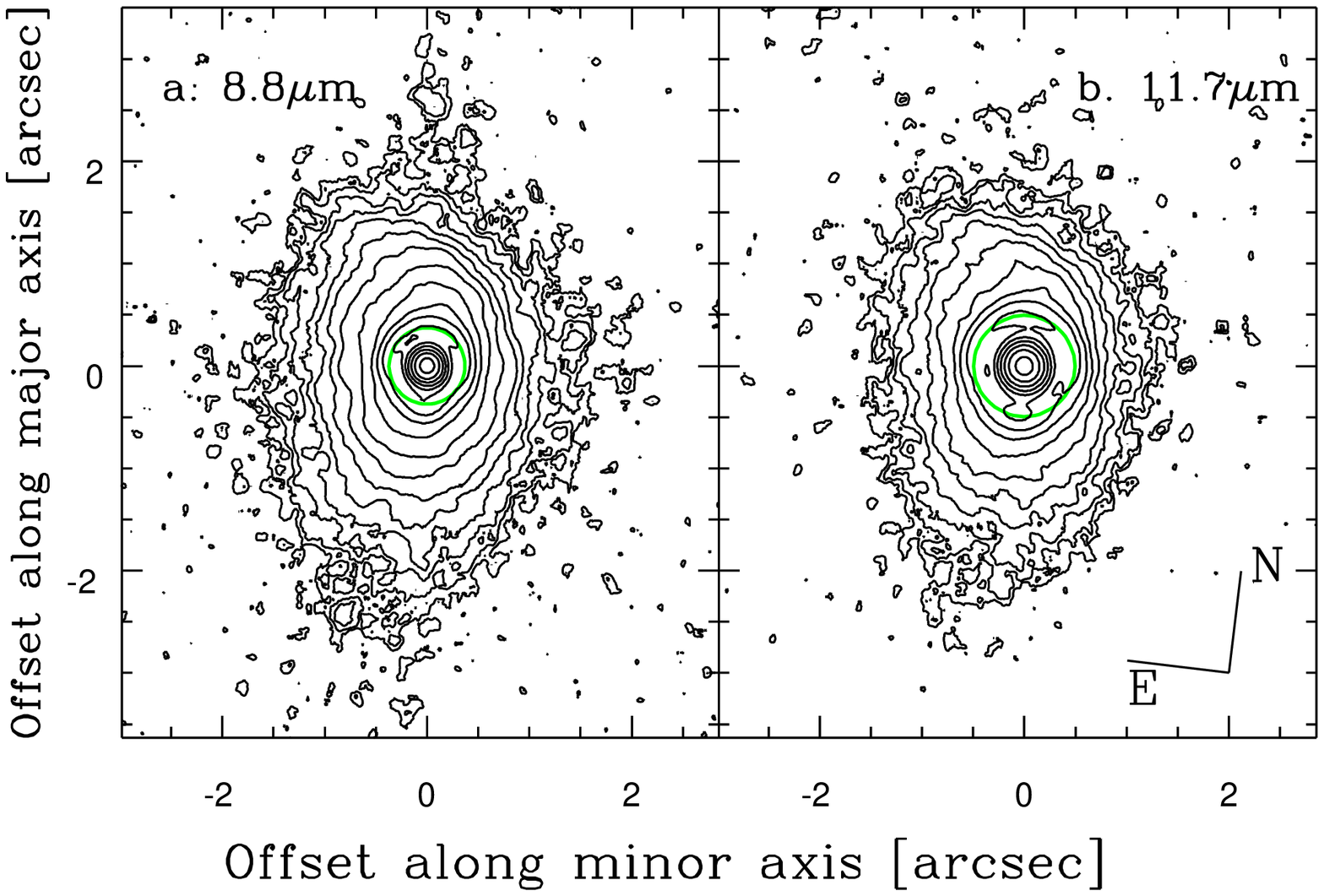}
\plotone{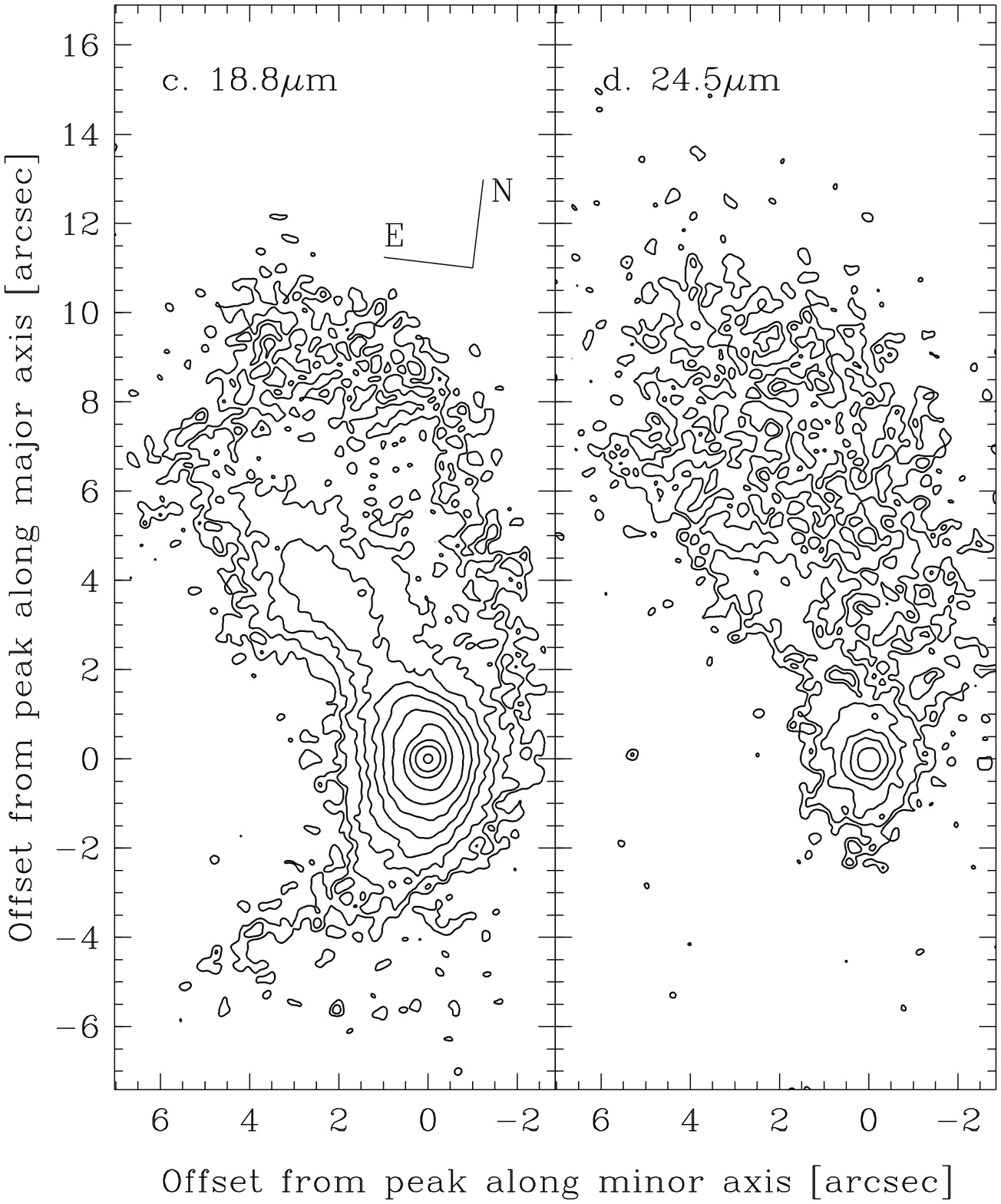}
\caption{
 Contour plot of the HD200775 images. 
The contour levels are 2$\sigma$ and 3$\sigma\times1.5^n$ ($n$=0, 1, 2, ... ),
where $\sigma$ is the sky noise level.
The origin of the coordinates of 1a-d is set at the peak of 
the emission and the axes are aligned with the major- and minor-axes of 
the disk. The green circles show positions of the brightest diffraction rings
(0.37$''$ and 0.49$''$ radius for 8.8\,$\mu$m and 11.7\,$\mu$m, respectively). 
(a) Contour plot of the 8.8\,$\mu$m image same as figure 1a (black), 
where $\sigma$ is 0.0163 [Jy/arcsec2]. 
(b) Contour plot of the 11.7\,$\mu$m image same as figure 1c (black), 
where $\sigma$ is 0.0188 [Jy/arcsec2]. 
(c) Contour plot of the 18.8\,$\mu$m image same as figure 1e, 
where $\sigma$ is 0.055 [Jy/arcsec2]. 
(d) Contour plot of the 24.5\,$\mu$m image same as figure 1g, 
where $\sigma$ is 0.189 [Jy/arcsec2].
}
\end{figure}

\begin{figure}
\epsscale{1.00}
\plotone{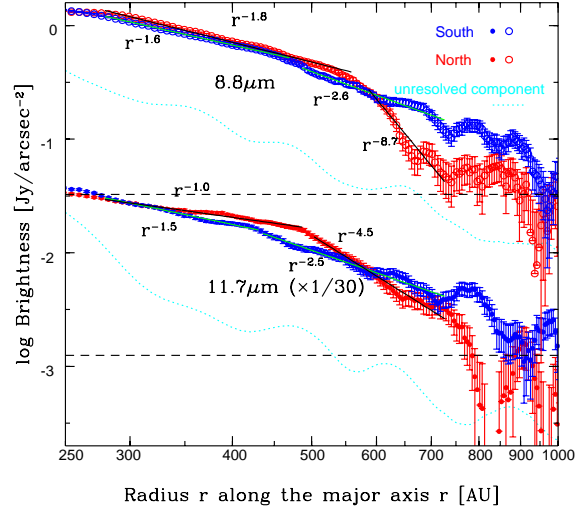}
\caption{
Brightness profile of the diffuse disk emission at 8.8 (open circles) 
and 11.7\,$\mu$m (filled circles; $\times$1/30). To be precise, 
the brightness at ($r$, $z(r)$) is plotted here, where $r$ is 
along PA=6.9degrees and $z(r)$ is based on the model fit results. 
The peak unresolved component is subtracted. 
Sky noise in the images is adopted as errors and 2\,$\sigma$ levels 
($\times$1/30 for 11.7$\mu$m) are shown by the black dashed lines. 
Power law functions ($r^{-p}$) are fitted to the profile and the 
results are shown with the green and black lines. 
For 11.7\,$\mu$m, the fitted power indices ($p$ of $r^{p}$) are 
-1.0 (280$<r<$490\,AU) and
-4.5 (500$<r<$730\,AU) in the north, and
-1.5 (280$<r<$430\,AU) and
-2.5 (440$<r<$730\,AU) in the south.
For 8.8\,$\mu$m, the indices are 
-1.8 (280$<r<$550\,AU) and
-8.7 (580$<r<$730\,AU) in the north, and
-1.6 (280$<r<$470\,AU) and
-2.6 (480$<r<$730\,AU) in the south.
The unresolved component profile at ($r$, $z(r)$) is estimated by 
the azimuthally averaged PSF profile (cyan). 
The diffuse emission of HD200775 is definitely brighter than the
skirt level of the PSF.
}
\end{figure}

\begin{figure}
\epsscale{1.00}
\plotone{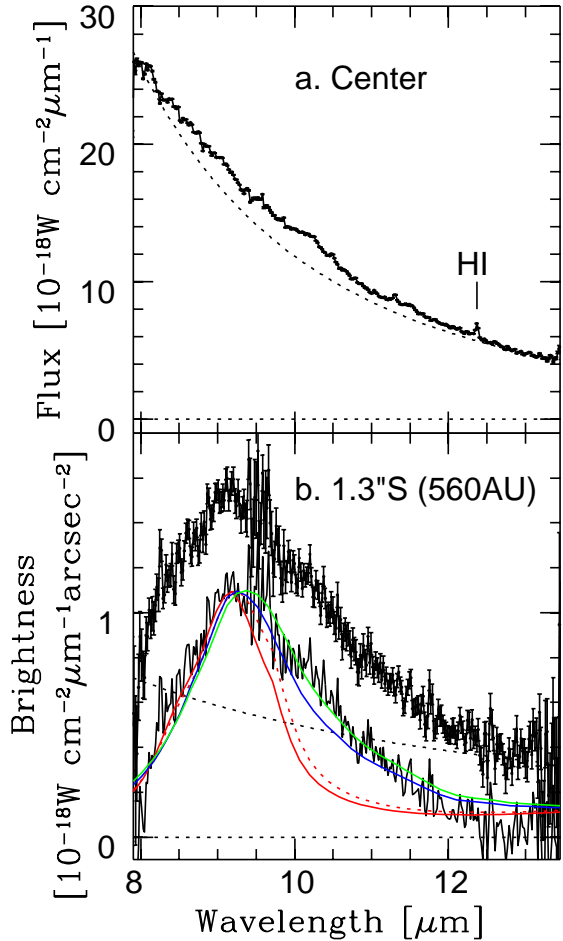}
\caption{
The 10\,$\mu$m region spectra from the COMICS spectroscopy. 
(a) At the unresolved peak source. 
(b) At the position 1.3$''$ south from the peak. 
We fitted the continuum emissions with power law spectra both 
at 8.0 and around 13.15\,$\mu$m individually (black dotted lines). 
In (a), it yields to 1600\,K blackbody emission. 
In (b), the continuum subtracted emission ($\times$1.15; black solid lines) 
and the normalized emission from amorphous silicate grains at 300\,K
based on J\"{a}ger et al. (2003) (colored lines) are shown. 
The red solid and dotted lines indicate those from Mg$_{0.7}$SiO$_{2.7}$ 
spherical grains with 0.1 and 1.0$\mu$m radii, respectively. 
The blue and green lines shows that of MgSiO$_{3}$ (pyroxene) 
spherical grains with 0.1 and 1.0$\mu$m radii, respectively.
}
\end{figure}

\begin{figure}
\epsscale{1.00}
\plotone{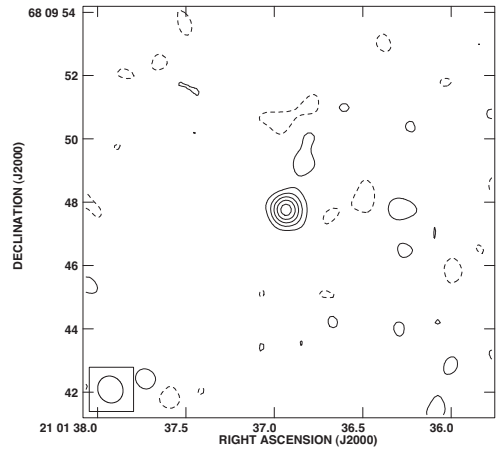}
\caption{
The 350.3\,GHz image obtained with SMA. The contours are 
6, 12, 18, 24, 30, 36, 42\,mJy/beam (solid lines) and -6\,mJy/beam 
(dashed lines). The lower-left panel shows the synthesized beam.
}
\end{figure}

\begin{figure}
\epsscale{1.00}
\plotone{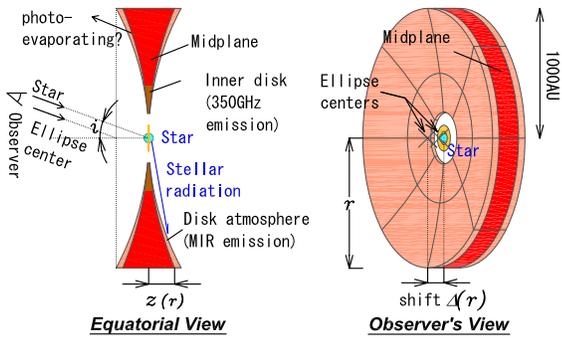}
\caption{
  Schematic view of the flared disk geometry. 
For an axisymmetric flared disk  where only the near side of the 
surface layer is seen with an inclination $i$, the observed isophotal 
contours become elliptical and the projected star position does not 
coincide with their centers. The shift ($\Delta$) increases as the major 
axis of the ellipse increases. The pink, brown (and red), and yellow 
areas indicate the optically thin part of the disk that emits 
in the mid-infrared, the optically thick midplane, and the 
innermost circumstellar disk, respectively. The disk surface is 
likely to photoevaporate outside the $r_{ph}$ ($\sim$70\,AU). 
The 18.8$\mu$m tail structure is not drawn here.
}
\end{figure}

\begin{figure}
\epsscale{0.60}
\plotone{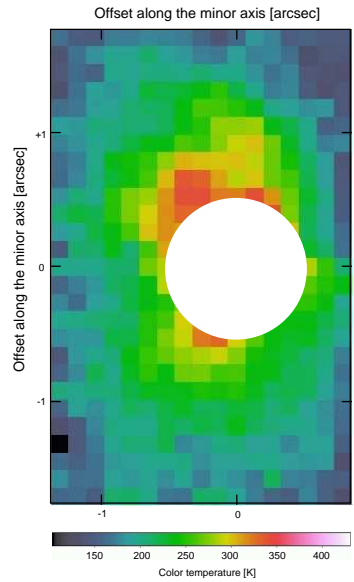}
\caption{
The distribution of the color temperature derived from the 11.7 
and 18.8\,$\mu$m brightness ratios.
The white circular area hides the area 
within the brightest Airy ring of the 11.7\,$\mu$m
band image (0.49$''$ radius) where the estimated value is significantly
affected by the diffraction pattern.
To increase the S/N ratio, 
10$\times$10 subpixels were averaged into one pixel with the original 
pixel size of the COMICS. 
}

\end{figure}
\begin{figure}
\epsscale{1.00}
\plotone{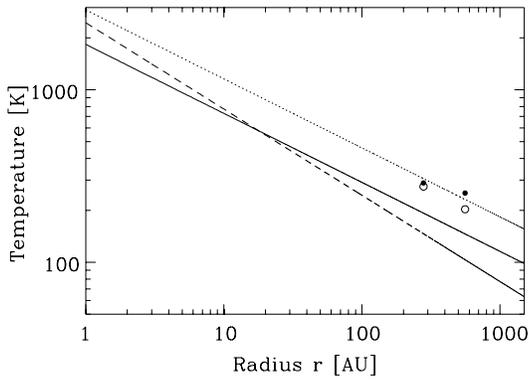}
\caption{
Temperature distribution of 
various grains expected around 6000\,L$_\odot$ star
for blackbody grains (dashed lines) and small grains that absorb
stellar emission perfectly and emit with emissivity of 
$a/\lambda$ in the infrared where $a$ is the grain radius.
The solid and dotted lines denote the 1 and 0.1\,$\mu$m radius grains,
respectively.
The color temperature estimated from the COMICS observations ($\S$4.2) were
plotted with open and filled circles for southern and northern side part
of the disk.
}
\end{figure}

\color{black}

\end{document}